\documentclass[12pt,preprint,longabstract]{aastex}

\usepackage{epsfig,graphicx,color}

\newcommand{\hii}{H\,{\scriptsize II}}

\newcommand{\av}{A$_{\rm V}$}

\shorttitle{Orion B}
\shortauthors{Schneider et al.}

\begin{document}

\title{What determines the density structure of molecular clouds ? \\
A case study of Orion B with {\sl Herschel}\footnote{Herschel is an ESA space
observatory with science instruments provided by European-led
Principal Investigator consortia and with important participation from
NASA.}}


\author{
N. Schneider\altaffilmark{1,2}, Ph. Andr\'e\altaffilmark{1}, V. K\"onyves\altaffilmark{1,3}, 
S. Bontemps\altaffilmark{2}, F. Motte\altaffilmark{1}, C. Federrath\altaffilmark{4,5}, 
D. Ward-Thompson\altaffilmark{6} and D. Arzoumanian\altaffilmark{1,3}, M. Benedettini\altaffilmark{7}, 
E. Bressert\altaffilmark{8}, P. Didelon\altaffilmark{1}, J. Di Francesco\altaffilmark{9}, 
M. Griffin\altaffilmark{10}, M. Hennemann\altaffilmark{1}, T. Hill\altaffilmark{1}, 
P. Palmeirim\altaffilmark{1}, S. Pezzuto\altaffilmark{7}, N. Peretto\altaffilmark{1},   
A. Roy\altaffilmark{1}, K.~L.~J. Rygl\altaffilmark{7}, L. Spinoglio\altaffilmark{7}, 
G. White\altaffilmark{11,12}   
}
\affil{
  $^1$IRFU/SAp CEA/DSM, Laboratoire AIM CNRS - Universit\'e Paris 
  Diderot, F-91191 Gif-sur-Yvette;
  $^2$OASU/LAB-UMR5804, CNRS, Universit\'e Bordeaux 1, F-33270 Floirac;
  $^3$IAS, CNRS/Universit\'e Paris-Sud 11, F-91405 Orsay;
  $^4$MoCA, School of Mathematical Sciences, Monash 
  University, Vic 3800, Australia;
  $^5$Inst. f\"ur Theor. Astrophysik, Universit\"at Heidelberg, D-69120 Heidelberg; 
  $^6$Jeremiah Horrocks Institute, UCLAN, Preston, Lancashire, PR1 2HE, UK; 
  $^7$IAPS-INAF, Fosso del Cavaliere 100, I-00133 Roma; 
  $^8$CSIRO Astronomy and Space Science, Epping, Australia; 
  $^9$NRCC, Herzberg Institute of Astrophysics,University of Victoria, Canada;  
  $^{10}$University School of Physics and Astronomy, Cardiff, UK; 
  $^{11}$Dep. of Physics \& Astronomy, The Open University, Milton Keynes MK7 6AA, UK;  
  $^{12}$RALSpace, Chilton, Didcot, Oxfordshire OX11, 0NL,UK
}
%\email{aastex-help@aas.org}
%\and
%\author{V. K\"onyves\altaffilmark{1,3}}
%\affil{ Monash Centre for Astrophysics (MoCA), School of Mathematical Sciences, Monash 
%  University, Vic 3800, Australia}


%\altaffiltext{1}{IRFU/SAp CEA/DSM, Laboratoire AIM CNRS - Universit\'e Paris 
%  Diderot, 91191 Gif-sur-Yvette, France}
%\altaffiltext{2}{OASU/LAB-UMR5804, CNRS, Universit\'e Bordeaux 1, 33270 Floirac, France}
%\altaffiltext{3}{IAS, CNRS/Universit\'e Paris-Syd 11, 91405 Orsay, France}
%\altaffiltext{4}{Monash Centre for Astrophysics (MoCA), School of Mathematical 
%Sciences, Monash University, Vic 3800, Australia}
%\altaffiltext{5}{a}


\begin{abstract}
  %Star formation in molecular clouds is governed by the interplay
  %between turbulence, gravity, and magnetic fields. 
  %, although their relative contributions are not settled. 
  A key parameter to the description of all star formation processes
  is the density structure of the gas. In this letter, we make use of
  probability distribution functions (PDFs) of {\sl Herschel} column
  density maps of Orion B, Aquila, and Polaris, obtained
  with the Herschel Gould Belt survey (HGBS).  We aim to understand
  which physical processes influence the PDF shape, and with which
  signatures.  
  %The Orion B PDFs are compared to those of the quiescent
  %Polaris region, and Aquila, another star-forming cloud.
%
  The PDFs of Orion B (Aquila) show a lognormal distribution for low
  column densities until \av\ $\sim$ 3 (6), and a power-law tail for
  high column densities, consistent with a $\rho\propto r^{-2}$ 
  profile for the equivalent spherical density distribution.
  % $\rho$ if we assume that the tail is only due
  %to self-gravity. 
  The PDF of Orion B is broadened by external compression due to the
  nearby OB stellar aggregates.  The PDF of a quiescent subregion of
  the non-star-forming Polaris cloud is nearly lognormal, indicating that
  supersonic turbulence governs the density distribution. But we also
  observe a deviation from the lognormal shape at \av\,$>$1 for a
  subregion in Polaris that includes a prominent filament.
  %which could be caused by statistical density fluctuations and not gravity. 
  We conclude that 
  (i) the point where the PDF deviates from the lognormal form does not trace a universal
  \av\,-threshold for star formation, 
  (ii) statistical density fluctuations, intermittency and magnetic fields can 
  cause excess from the lognormal PDF at an early cloud formation stage,
  (iii) core formation and/or global collapse of
  filaments and a non-isothermal gas distribution lead to a power-law tail,
  and (iv) external compression broadens the column density PDF,
  consistent with numerical simulations.
\end{abstract}

%\keywords{ISM: clouds, structure}



\section{Introduction} \label{intro}

%Density structure of clouds and the different components 
The star formation process represents a dramatic transformation of a
molecular cloud in time and space where the main governing elements 
are turbulence, gravity, and magnetic fields.  The spatial structure
of clouds, now impressively revealed by {\sl Herschel} imaging
observations in the far-infrared (e.g., Andr\'e et al. \cite{andre2010},
Motte et al. \cite{motte2010}, Molinari et al. \cite{molinari2010}),
is very inhomogeneous and dominated by filaments (Arzoumanian et
al. \cite{doris2011}, Schneider et al. \cite{schneider2012}). 
%The density in molecular clouds ranges between diffuse interclump
%gas (a few particles cm$^{-3}$) and star-forming
%cores ($>$10$^4$ cm$^{-3}$). 
It is only with {\sl Herschel} space observatory (Pilbratt et al.
\cite{pilbratt2010}) observations that diffuse to dense gas is now
traced at high angular resolution (typically 18$''$). Combining PACS
(Poglitsch et al.  \cite{poglitsch2010}) and SPIRE (Griffin et al.
\cite{griffin2010}) data provides column density maps that are
superior to those obtained from extinction using near-IR data
(Lombardi et al.  \cite{lombardi2006}, Kainulainen et al.
\cite{kai2009}, Froebrich \& Rowles \cite{froebrich2010}, Schneider et
al. \cite{schneider2011}) that have angular resolutions of
$\approx$2$'$ and suffer from saturation at visual extinctions \av\
above $\approx$25.

%PDFs 
%With the {\sl Herschel} data in hand, it is now possbile to study the
%(column)-density structure of molecular clouds in much greater detail.
Here, we aim to disentangle the relative contributions of turbulence,
gravity, and external compression that influence the density structure
of a molecular cloud. A useful analysis technique is to use
probability distribution functions (PDFs) of the column density, which
characterizes the fraction of gas with a column density $N$ in the
range [$N$, $N$+$\Delta N$] (e.g., Federrath et al.  \cite{fed2010}).
Extinction maps (see above) have shown that molecular clouds can have
a lognormal PDF for low column densities, and either a power-law tail
or more complex shapes for higher column densities. Isothermal,
hydrodynamic simulations including turbulence and gravity (e.g.,
Klessen et al. \cite{klessen2000}) have shown that gravitational
collapse induces a power-law tail in the PDF at high densities. More
recent studies (Kritsuk et al. \cite{kritsuk2011}, Federrath \&
Klessen \cite{fed2013} and references therein)
%(Federrath et al. \cite{fed2008},
%Ballesteros-Paredes et al. \cite{ball2011}, Cho \& Kim \cite{cho2011},
%Kritsuk et al. \cite{kritsuk2012}, Collins et al. \cite{collins2012})
have investigated which parameters influence the shape of the PDF.
Following these studies, fitting the slope of the high-density tail of
the PDF allows us to determine the exponent $\alpha$ of an equivalent
spherical density distribution $\rho(r)=\rho_0 \, (r/r_0)^{-\alpha}$.

\begin{figure*}[ht]
\begin{center} 
%\hspace{-0.5cm}
\includegraphics [width=13cm, angle={0}]{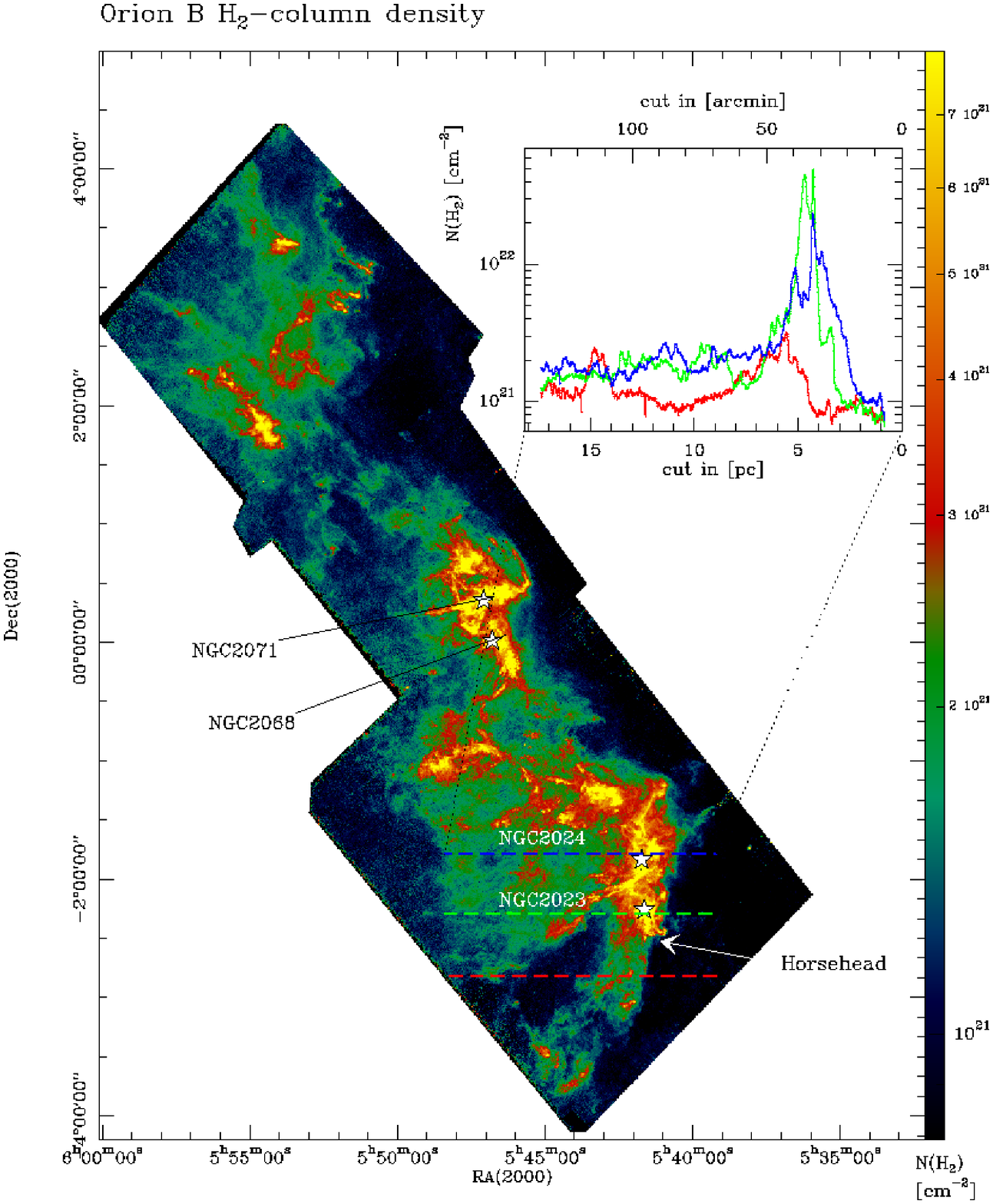}
\caption [] {H$_2$-column density map at 18$''$
  angular resolution of Orion B obtained from {\sl Herschel} data.
  Known \hii\ regions are labeled.  The panel inside the image shows cuts (color-coded
  in blue, green, and red) in H$_2$-column density of the NGC2023/24
  region at constant declination. These cuts are indicated in the image.}
\label{orionb}
\end{center} 
\end{figure*}

\begin{figure*}[ht]
\begin{center} 
\includegraphics [width=13cm, angle={0}]{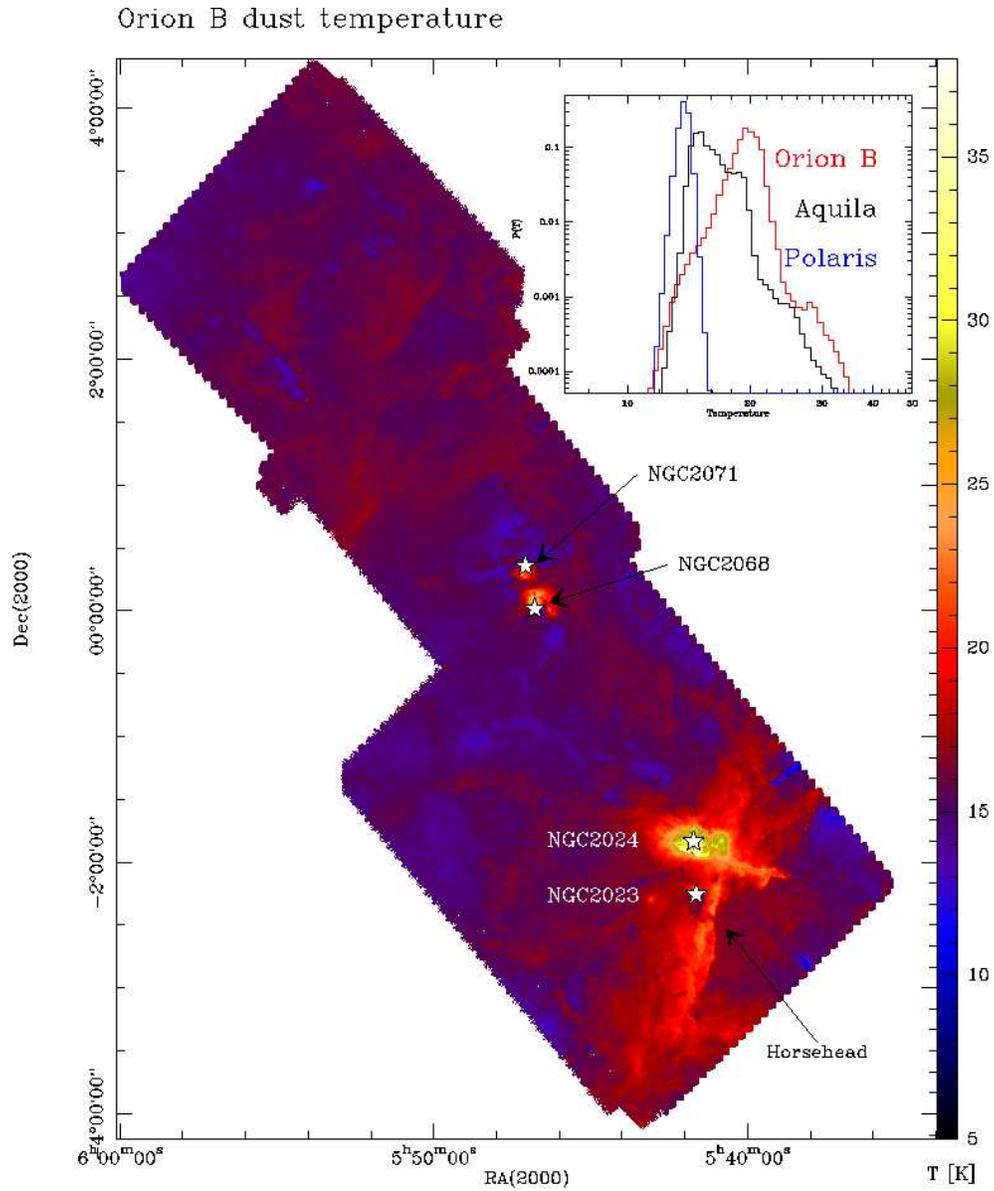}
\caption [] {Dust temperature map at 18$''$
  angular resolution of Orion B obtained from {\sl Herschel} data.
  The panel inside the image shows the temperature PDFs of Orion B, Aquila, and 
Polaris (the whole region, not seperated by subregions).}
\label{orionb-temp}
\end{center} 
\end{figure*}

% Orion B  distance 390 pc for NGC2071 and 415 pc for NGC2024
In this study we make use of {\sl Herschel}-derived column density
PDFs of the \object{Orion B} molecular cloud, a template region for
studies of low- to high-mass star formation (Lada et
al. \cite{lada1991}). Orion B is amongst the nearest (distance
$\sim$400 pc; Gibb \cite{gibb2008}) giant molecular cloud (GMC) complexes,  
with a mass of around 10$^5$ M$_\odot$, and hosts several OB-clusters
(NGC2023/24, NGC2068/71). Orion B is located within the
H$\alpha$-shell 'Barnard's Loop' and diverse OB stellar aggregates impact
the cloud from the west with radiation and stellar winds.
%This compressive energy injection has been proposed to have 
%triggered star formation in Orion B (e.g., Bally et al.
%\cite{bally1987}).  
To understand better what governs the density
structure and its link to star formation, we compare the Orion B PDFs
to those obtained with the HGBS for a quiescent cloud (Polaris) and
a star-forming region (Aquila). 

\begin{figure*}[ht]
\begin{center}
\includegraphics[angle=0,width=10cm]{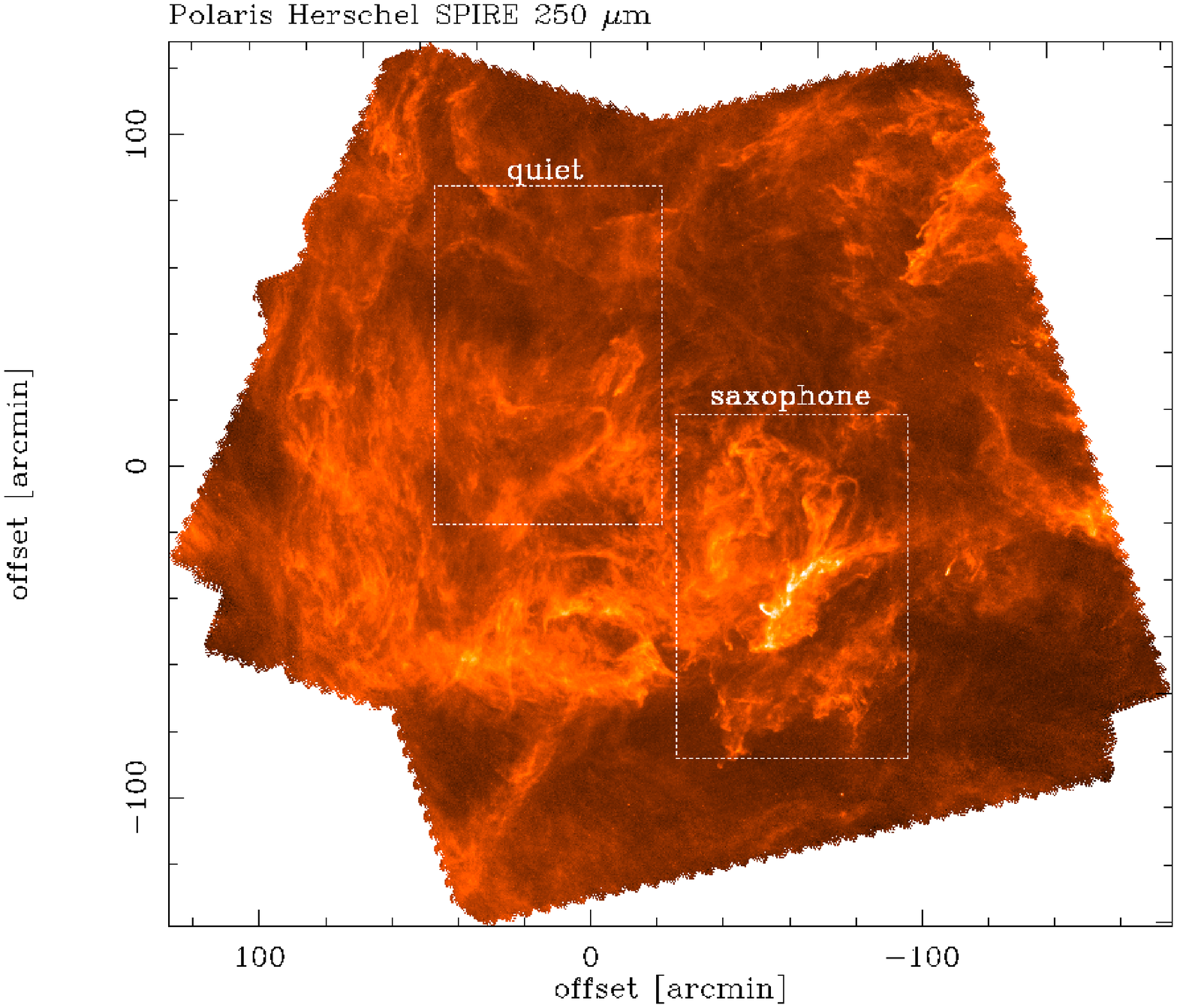}
\includegraphics[angle=0,width=10cm]{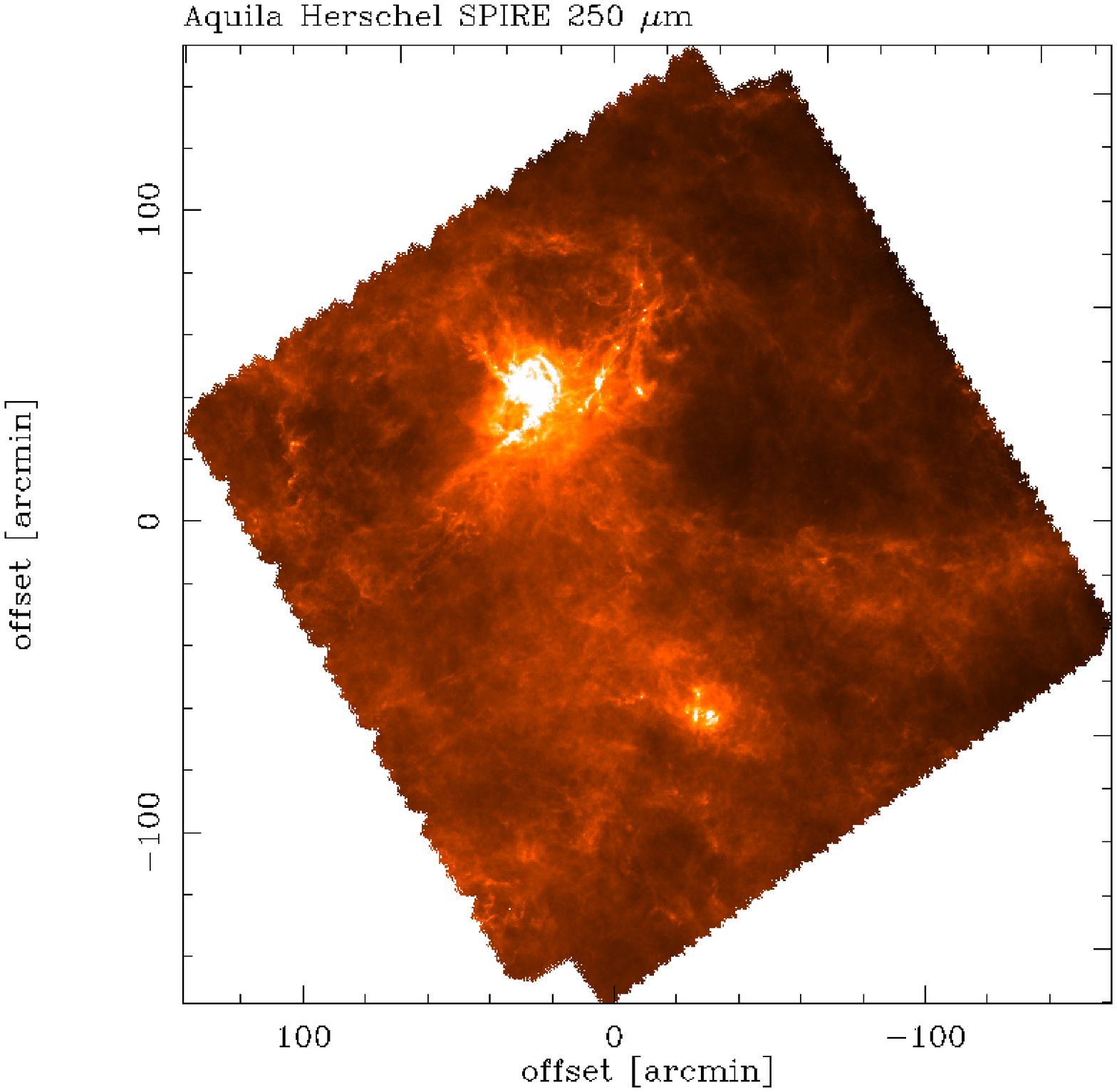}
\end{center}
\caption [] {{\sl Herschel} SPIRE map at 250 $\mu$m of the Polaris  and 
Aquila fields. The two regions for which we show the corresponding column 
density PDFs for Polaris are indicated by dashed white lines. These are the 'saxophone', a
region containing a filament with high column density but lacking star formation 
%(Andr\'e et al. \cite{andre2010}, Ward-Thompson et al.\cite{derek2010}), 
and a subregion further north-east. 
%(Miville-Desch\^enes et al.  \cite{miville2010},Men'shchikov et al. \cite{sascha2010}).
}
\label{polaris}
\end{figure*}

\section{Observations}

Orion B, Aquila, and Polaris were observed with the PACS and
SPIRE instruments onboard {\it Herschel} as part of the Herschel Gould
Belt survey (HGBS, Andr\'e et al. \cite{andre2010}) in parallel mode
with a scanning speed of 60$''$/sec and two orthogonal coverages. The
Orion B data were obtained on 2010 Sep 29 and 2011 Mar 13. For details
on Polaris see Men'shchikov et al. \cite{sascha2010},
Miville-Desch\^enes et al.  \cite{miville2010}, Ward-Thompson et
al. \cite{derek2010}, and K\"onyves et al. \cite{vera2010}, Bontemps
et al. \cite{bontemps2010} for Aquila.
%
%The PACS 70, 160 $\mu$m and SPIRE 250, 350, 500 $\mu$m data are
%publicly available in the {\sl Herschel} Science Archive.
%
The angular resolutions at 160 $\mu$m (PACS), 250 $\mu$m, 350 $\mu$m, and
500 $\mu$m (all SPIRE), are $\sim$12$''$, $\sim$18$''$, $\sim$25$''$,
and $\sim$36ß$''$, respectively. The SPIRE data were reduced with HIPE
version 7.1956, including a destriper-module with a polynomial
baseline of zeroth order.  Both scan directions were then combined using
the `naive-mapper', i.e., a simple averaging algorithm.  The PACS data
were reduced using HIPE 6.0.2106. In addition to the standard data
reduction steps, non-linearity correction was applied on the 160
$\mu$m signal, which affects only the bright ($>$ 10 Jy/pixel) regime.
The level1 data were then combined into a map with Scanamorphos v10
(Roussel \cite{roussel2012}).

Column density and dust temperature maps were determined from a
modified blackbody fit to the wavelengths 160 to 500 $\mu$m (see,
e.g., K\"onyves et al. \cite{vera2010}).  We recovered the {\it
Herschel} zero-flux levels of the Orion B field for each wavelength
with Planck data (Bernard et al. \cite{bernard2010}). For the region
covered by both PACS and SPIRE simultaneously, we fixed the specific
dust opacity per unit mass (dust+gas) approximated by the power law
$\kappa_\nu \, = \,0.1 \, (\nu/1000 GHz)^\beta$ cm$^{2}$/g and
$\beta$=2 (cf. Hildebrand \cite{hildebrand1983}), took a mean
molecular weight per hydrogen molecule of 2.8, and left the dust
temperature and column density as free parameters.  As an improvement
to this procedure, we applied the technique described in Palmeirim et
al.  (\cite{pedro2013}) that uses the flux information of the 500
$\mu$m map but with the help of a multi-resolution decomposition, the
angular resolution of the final maps is higher, i.e., that of the 250
$\mu$m data at 18$''$.  To test the robustness of the derived
high-resolution map of Orion B, we constructed ratio maps between the
18$''$-resolution column density map -- smoothed to the
common-resolution of the 500 $\mu$m data (36$''$)-- and the originally
36$''$-resolution maps. This ratio map has a mean value of 1.0 and a
standard deviation of 0.03.  The two column density maps agree within
15\%. In addition, we investigated the effect of increasing opacity
for high column densities (Roy et al. (\cite{roy2013}) on the PDF and
found that its dispersion decreases ($\sim$10--20\%) and can provoke a
steeper slope of the power-law tail for high densities.

Because the density structure of molecular clouds depends on how
energy is injected into a cloud (spiral density waves, expanding
supernovae shells, \hii-regions, or gravitational contraction), we
determine the hydrodynamic Mach-number $\mathcal{M}$ that
characterizes to first order the influence of turbulence. Stronger
isothermal, non-magnetized supersonic turbulence leads to a higher
Mach number and thus stronger local density enhancements.  In
contrast, magnetic fields smooth out density variations (Molina et al.
\cite{molina2012}). $\mathcal{M}$ can be derived from observations of
the full width at half maximum (FWHM in [km s$^{-1}$]) of a molecular
line, and the sound speed $c_s$=0.188$\sqrt{T_{kin}/10 K}$ with the
kinetic temperature T$_{kin}$.

\begin{equation}
\mathcal{M}=(\sqrt{3}\, FWHM)/(c_{s}\, \sqrt{8 \ln2})   
\end{equation}

 If the LTE assumption is valid, T$_{kin} \approx$T$_{ex}$, with
 T$_{ex}= 5.53 [\ln (5.53/T_{mb})+1)]^{-1}$ for the optically thick
 $^{12}$CO 1$\to$0 line. If gas and dust are well mixed, the
 temperature should also correspond to the dust temperature derived
 from {\sl Herschel}.  We emphasize, however, that the determination
 of the Mach-number remains rather uncertain (error $\sim$30--40\%)
 and mainly gives a {\sl tendency}.

\begin{figure*}[ht]
\begin{center}
\includegraphics[angle=90,totalheight=5.5cm]{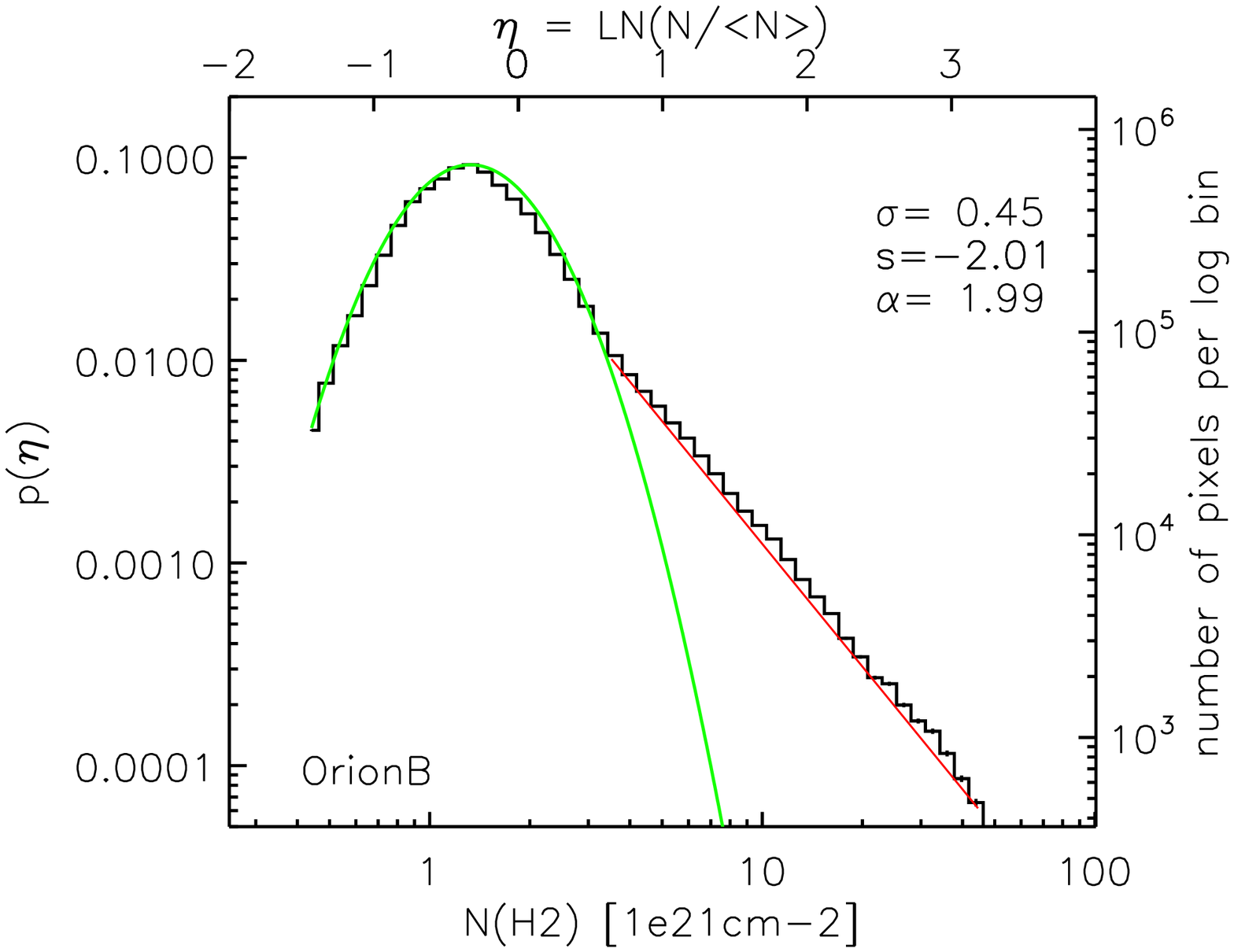}
\hspace{1cm}\includegraphics[angle=90,totalheight=5.5cm]{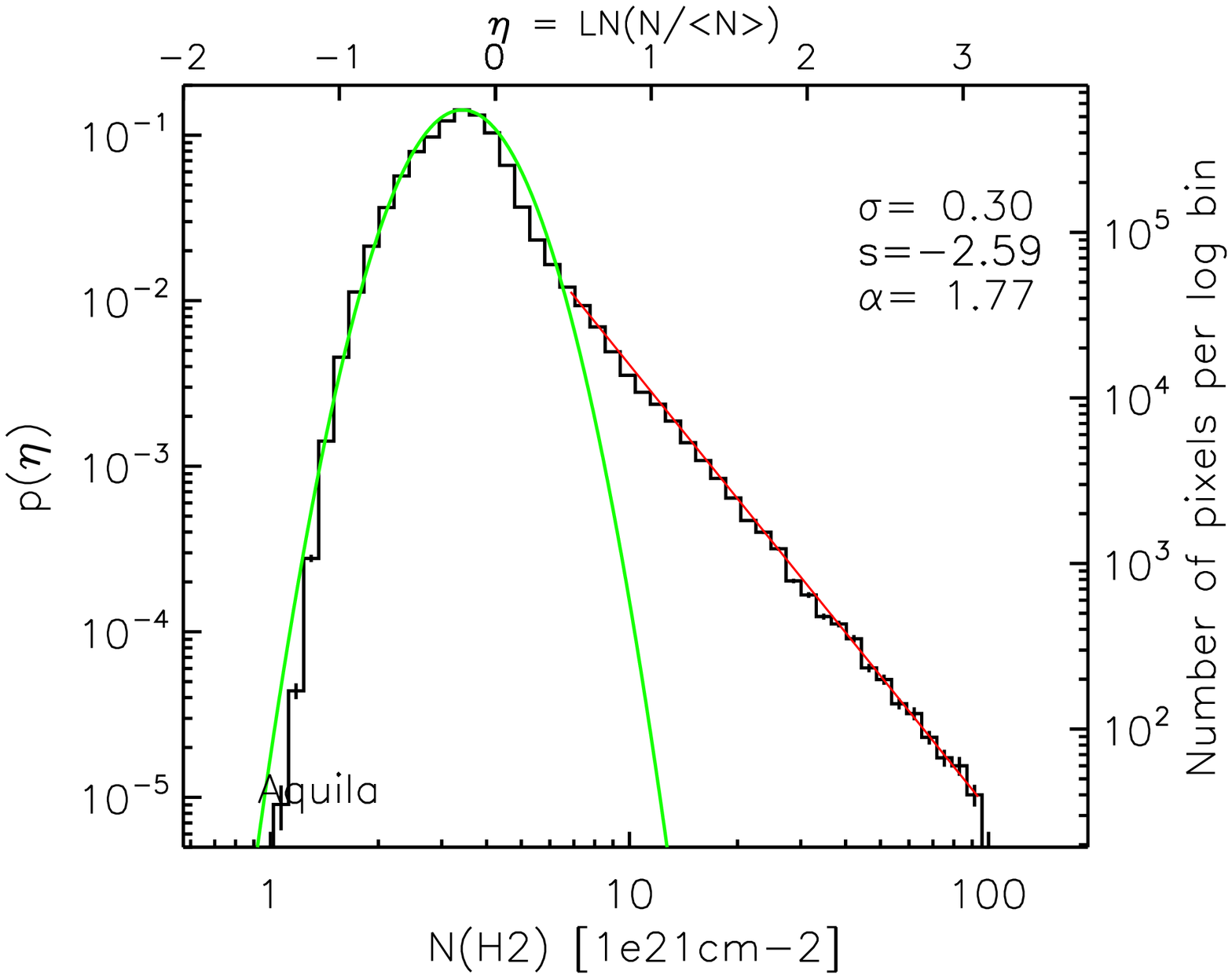}
\end{center}
\begin{center}
\includegraphics[angle=90,totalheight=5.5cm]{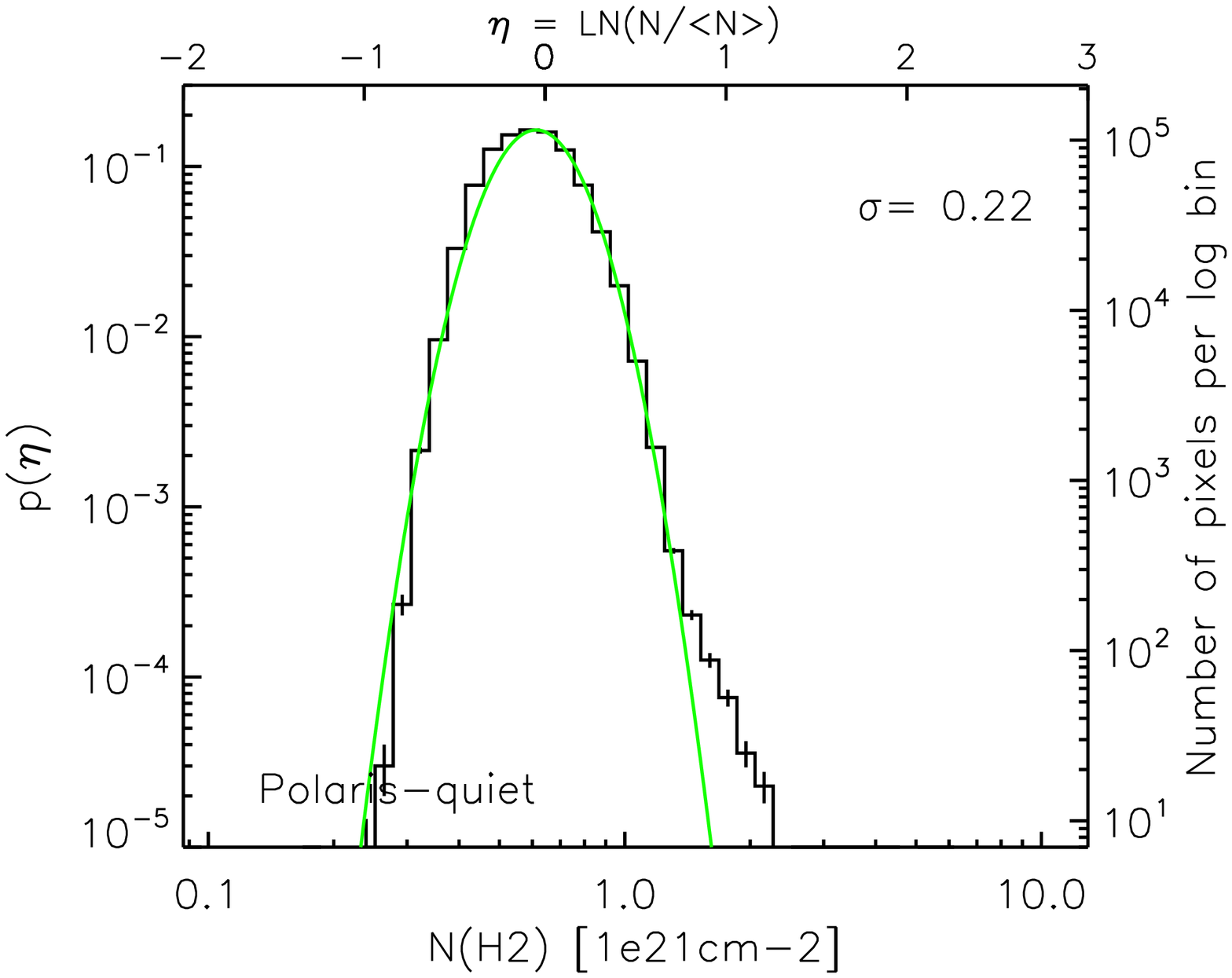}
\hspace{1cm}\includegraphics[angle=90,totalheight=5.5cm]{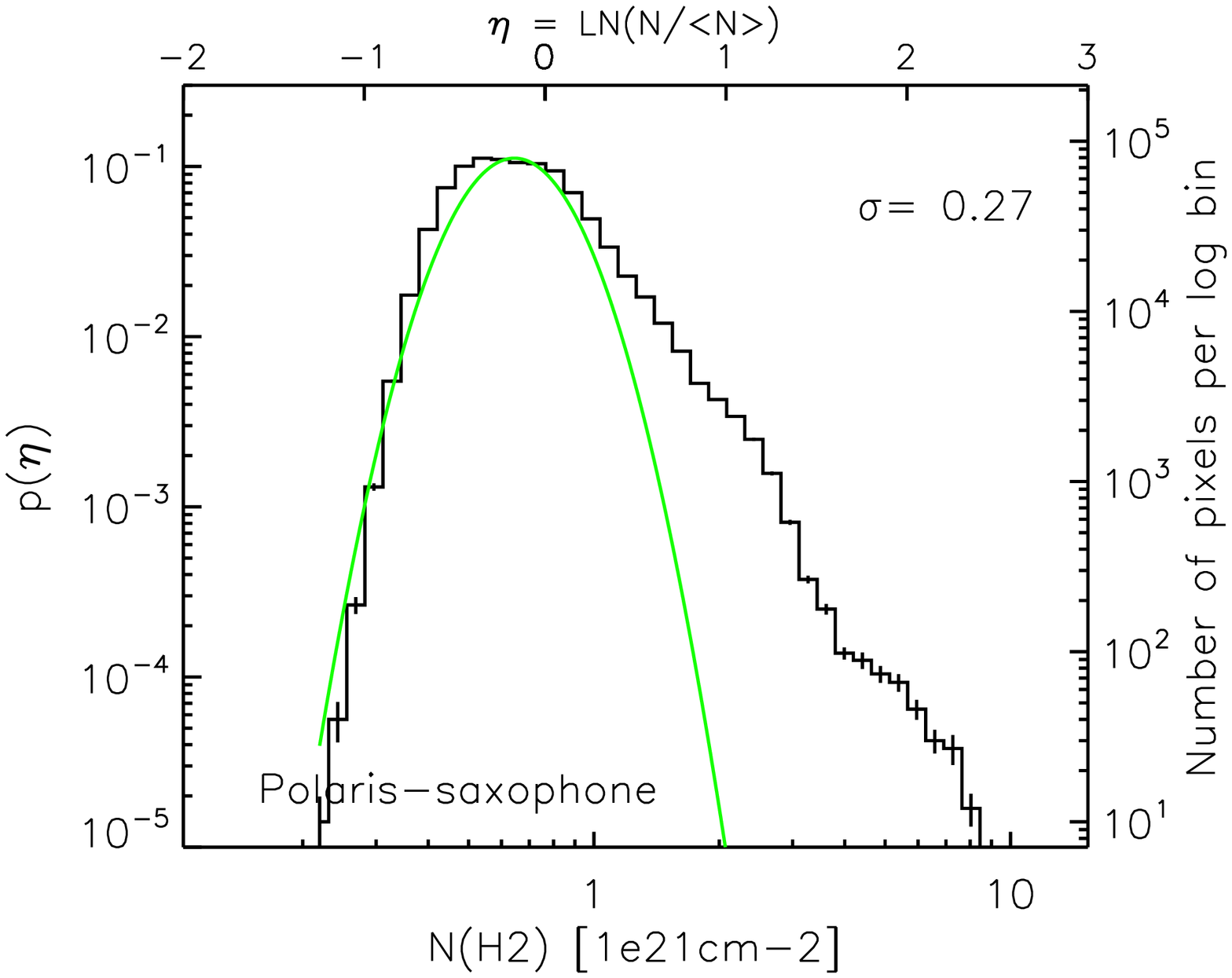}
\end{center}
\caption [] {Probability distribution functions of column density for
  Orion B and Aquila (top) and Polaris (bottom) at an angular
  resolution of 18$''$ (a similar version of the Aquila PDF was
  previously shown in Andr\'e et al. \cite{andre2011}).  The left
  y-axis is the PDF as a normalized probability $p(\eta)$ while the
  right y-axis indicates the number of pixels per logarithmic bin.
  Note that due to the large number of pixels (6$''$ grid), the
  error-bars calculated using Poisson-statistics are very small. The
  PDFs do not change using a lower sampled grid (i.e. at the
  resolution of 18$''$ or 36$''$). The lower x-axis gives the column
  density N(H$_2$) in units of 10$^{21}$ cm$^{-2}$ (corresponding
  approximately to the \av\, in magnitudes using the Bohlin-factor).
  The upper x-axis is the dimensionless parameter $\eta=\ln(N/\langle
  N \rangle)$. The green curve indicates the fitted lognormal PDF and
  the red line the power law fit to the high-density tail. The width
  of the fitted lognormal PDF ($\sigma_{\eta}$), the powerx-law slope
  index $s$, and the exponent $\alpha$ of the equivalent spherical
  density profile ($\rho \propto r^{-\alpha}$) are given in each
  panel. Note that the variation of opacity with density causes a
  systematic error on the PDF which is larger than the statistical
  error. We estimate the error on $\sigma$ and $\alpha$ to be
  typically 10--20\%.}
\label{pdf1}
\end{figure*}
 
\section{The column density structure of Orion B} \label{results}

\subsection{Column density maps} \label{skl} 

% Page 9 Handbook, chapter Orion B  -> NGC2024 'ridge'
%
The column density map of Orion B (Fig. ~\ref{orionb}) is dominated by
the two active star-forming clumps NGC2023/24 and NGC2068/71 (Buckle
et al. 2010) with very high local column densities N(H$_2$) up to
a few 10$^{22}$ cm$^{-2}$ and high dust temperatures of up to 35 K
(Fig.~\ref{orionb-temp}) due to the \hii-regions.  These two dense
ridges are outlined by a column density level of
$\sim$3--4$\times$10$^{21}$ cm$^{-2}$ and stand out in an extended
cloud with a typical column density of 1--2$\times$10$^{21}$
cm$^{-2}$.  In contrast, the northern-eastern part of Orion B is
colder and less active.  A sharp cut-off in column density at the
western border of NGC2023/2024 and the Horesehead nebula is seen in column
density cuts at constant declination (Fig.~\ref{orionb}) which was known
from CO data (Wilson et al.  \cite{wilson2005}).  From west to east,
we first observe a strong increase of column density on a few pc scale
for NGC2023/2024, and a weaker but clearly visible increase for the
southern region. East of the peaks, the column density decreases to a
level of 1 to 2$\times$10$^{21}$ cm$^{-2}$ which is higher than the
values at the western border. Such a profile 
was also seen in the Pipe nebula (Peretto et al. \cite{peretto2012}) where
the authors proposed a large-scale compression by the winds of the Sco
OB2 association likely caused this sharp edge. A similar process may
be at work for Orion B because the western part of the cloud is
exposed to various OB aggregates (OB1b--d).

\subsection{Probability distribution functions of  column density} \label{prob} 

The distribution of number of pixels vs. column density for Orion B,
Aquila, and Polaris are displayed in Fig.~\ref{pdf1}. We will use the 
term probability distribution function (PDF) and the notation
p($\eta$) though the pure pixel distribution is not
strictly a PDF which is defined for a lognormal distribution as 
\begin{equation} 
p(\eta)d\eta=(2\,\pi\,\sigma^2_{\eta})^{-0.5} \,
\exp[-(\eta-\mu)^2/(2\sigma^2_{\eta})]\,d\eta
\end{equation} 
with $\eta=\ln(N/\langle N \rangle)$ and $\sigma_{\eta}$ as the
dimensionless dispersion of the logarithmic field, and $\mu$ the mean.
The normalization allows a direct comparison between clouds of
different column density, and $\sigma_{\eta}$ is a measure for the
density variation in a turbulent medium.  We determine $\sigma_{\eta}$
with a fit to the assumed lognormal low-density part of the PDF and
the slope with the index $s$ from a power-law fit with $p(\eta) = p_0
(\eta/\eta_0)^s$ to the high-density part.  The results of the 
PDF fit are given together with the Mach-number determination in Table 1. 

\begin{table*}[] 
\begin{center}
\caption{Temperature regime, Mach-number, and PDF fit results for Orion B, Aquila, and Polaris }
\begin{tabular}{lcccccccc}
\tableline\tableline
                       &              &            &               &              &                 &     &  &\\
                       & $T_{ex}$(CO) & $T_{dust}$ & $\langle T_{ex} \rangle$(CO) & $\langle T_{dust} \rangle$ & $\Delta{\rm v}$ & $\mathcal{M}$ & $\sigma_{\eta}$ & $\alpha$\\
                       &   [K]        & [K]       & [K]            &  [K]         &     [km/s]      &     &     & \\
 Cloud                 &   (1)        & (2)       & (3)            &  (4)         &     (5)         & (6) & (7) & (8) \\
\tableline
{\bf OrionB}           &  5--70      & 5--45      &  20            &  16          &  $\sim$3        & $\sim$8 & 0.45 & 1.99 \\
{\bf Aquila}           &   ...       & 9-40       &  20            &  19          &  $\sim$2.2      & $\sim$6 & 0.30 & 1.77 \\
{\bf Polaris-quiet}    &   ...       & 12-15      &  10            &  13          &  $\sim$1        & $\sim$3 & 0.22 &  - \\
{\bf Polaris-saxophone}& $\sim$10-15 & 11-14      &  12            &  13          &  $\sim$2        & $\sim$7 & 0.27 &  - \\
\tableline
\end{tabular}
\tablenotetext{}{(1) Observed excitation temperature range from $^{12}$CO 1$\to$0 data 
(Orion B: Buckle et al. \cite{buckle2010}, Aquila: Zeilik et al. \cite{zeilik1978};   
Polaris: Bensch et al. \cite{bensch2003}, Shimoikura et al. \cite{shi2012}). 
No large-scale CO data is availabe for Polaris and Aquila).} 
\tablenotetext{}{(2) Observed dust temperature range from {\sl Herschel} data.}
\tablenotetext{}{(3) and (4) average temperature.}
\tablenotetext{}{(5) line width from $^{12}$CO 1$\to$0.}
\tablenotetext{}{(6) sonic Mach-number from average temperature and CO-line width.}
\tablenotetext{}{(7) dispersion of the PDF.}
\tablenotetext{}{(8) exponent of the spherical density profile.}
\end{center}
\end{table*}

The Orion B and Aquila PDFs show a well-defined lognormal part for low
column densities and a clear power-law tail at higher column
densities, starting at an extinction\footnote{For better comparison to
the literature values, we use the visual extinction value derived from
the column density adopting the conversion formula
N(H$_2$)/\av=0.94$\times$10$^{21}$ cm$^{-2}$ mag$^{-1}$ (Bohlin et
al. \cite{bohlin1978}).} \av\, around 3 for Orion B and 6 for
Aquila. To first order, the PDFs only differ in their width
($\sigma_{\eta}$=0.45 for Orion B and 0.3 for Aquila).  The PDFs of
two subregions (Fig.~\ref{polaris}) in the Polaris cirrus cloud
(Falgarone et al. \cite{falgarone1998}) are more narrow with
$\sigma$=0.22 and 0.27, respectively. The PDF of the quiescent region
is almost perfectly lognormal, however, above \av $\sim$1.5 we observe
a slight excess which ist most likely a resolution effect but may
possibly result from a physical process (see Sec. 3.3).  A clear
deviation from the lognormal shape is found for the 'saxophone'
filament.  Interestingly, the excess for \av\,$>$1 has not the form of
a power-law tail. Note that all PDF features (shape, width etc.) do
not depend on the angular resolution (18$''$ vs. 36$''$) or pixel
number (our statistic here is high because the images are on a 6$''$
grid).  To quantify better the deviation of the PDF from lognormal, we
determined the higher moments\footnote{$\mathcal{S}={\frac{1}
{\sigma^3}} \int_{-\infty}^{\infty} d\eta\, p(\eta) [\eta-\langle \eta
\rangle]^3$ and $\mathcal{K}={\frac{1}{\sigma^4}}
\int_{-\infty}^{\infty} d\eta\, p(\eta) [\eta-\langle \eta \rangle]^4$
(see, e.g., Federrath et al.  2010)} skewness $\mathcal{S}$ and
kurtosis $\mathcal{K}$.  The skewness, describing the asymmetry of the
distribution, is positive for Orion B, Aquila, and Polaris-saxophone
($\mathcal{S}$=1.17, 1.24, and 0.49, error typically 0.05), implying
an excess at higher column densities, and is $\mathcal{S}$=0.19
($\mathcal{K}$=3.0) for Polaris-quiet, confirming the nearly 
lognormal form of its PDF.  The Orion B and Aquila PDFs have much
higher values for the kurtosis ($\mathcal{K}$=6.8 and 7.9) that arise
from pronounced wings. The Polaris-saxophone region has also a (high)
value of $\mathcal{K}$=4.4, indicating an excess at high column
densities.

%A normal distribution is characterized by a
%kurtosis of three. A distribution closer to a box profile shows lower
%values, distributions with broads wings have a higher kurtosis.

From theory, a purely lognormal distribution is only expected if the
cloud structure is shaped by supersonic, isothermal turbulence while
deviations in the form of a power law for high column densities are
predicted for self-gravitating clouds (Klessen et al.
\cite{klessen2000}, Ballesteros-Paredes et al.  \cite{ball2011}).
The concept of isothermality does not fully apply to all clouds, as
can be seen in the temperature PDFs in Fig.~\ref{orionb-temp}. While
Polaris can be considered as a nearly isothermal gas phase, Orion B
and Aquila show a more complex temperature distribution over a larger
range.  Froebrich \& Rowles \cite{froebrich2010} argued that the
\av-value where the transition of the PDF must take place is always
around 6, and defined this value as a threshold for star formation.
Kainulainen et al. \cite{kai2011} found values between
\av=2--5 and proposed a scenario in which this \av-range marks a
transition between dense clumps and cores and a more diffuse
interclump medium. The PDFs of Orion B and Aquila shown in
Fig.~\ref{pdf1} clearly show that the transition from lognormal to
power law is not universal but varies between clouds (\av\,$\sim$3 and
6, respectively). This behaviour suggests that the \av\,-transition
value neither represents a universal threshold in star formation nor a
phase transition (unless the density of the clumps and the interclump
medium strongly varies from cloud to cloud). A similar result was
obtained in the study of Schneider et al. (2013), presenting a large
sample of PDFs from low- to high-mass star-forming clouds.  Deviations
of the PDF from lognormal are more likely a function of cloud
parameters, in particular the virial parameter, the dominant forcing
mode\footnote{{\sl Compressive} modes on large to small scales are
  generated by galactic spiral shocks, expanding supernova shells and
  \hii\, regions, gravitational contraction, and outflows. {\sl
    Solenoidal} forcing arises from galactic rotation and
  magneto-rotational instabilities.}, and the Mach number as shown in
models (Federrath \& Klessen 2013) and observations (Schneider et al.
2013).

From the slope of the power-law tail for Orion B and Aquila, the
exponent of an equivalent spherical density distribution
$\rho(r)\propto r^{-\alpha}$ is determined to be $\alpha$=1.99 (1.77)
for Orion B (Aquila), conforming with results typically obtained for
individual collapsing cores. The high-density tail, however, cannot be
explained by the core population alone because it does not provide
sufficient mass (K\"onyves et al. \cite{vera2010}).  Moreover, it is
probably also caused by global gravitational collapse of larger
spatial areas like filaments and ridges (see, e.g., Schneider et
al. \cite{schneider2010}, Hill et al. \cite{hill2011}, Palmeirim et
al. \cite{pedro2013} for observational examples).  It was shown that
non-isothermal flows can also cause power-law tails (Passot
\&Vazquez-Semadeni 1998) so that this process may influence the shape
of the PDF as well. Indeed, in regions with significant temperature
variations, $\alpha$ can reach values larger than free-fall
($\alpha$=2.4 for the high-mass star-forming cloud NGC6334, Russeil et
al. \cite{russeil2013}), possibly pointing towards a scenario in which
heating/cooling processes become important.

\subsection{Comparison to models} \label{model} 

We compare our observational PDFs with those obtained from
hydrodynamic simulations (Federrath \& Klessen 2013), including
gravity, magnetic fields, and different turbulent states 
($\mathcal{M}$=2--50, star formation efficiencies from 0 to 20\%, and
different forcing modes). In addition, we determine the
  'b-parameter' ($\sigma_s^2=f^2\, \sigma_{\eta}^2= \ln{(1+b^2\,\mathcal{M}^2)}$, e.g., 
    Federrath et al. 2010, Burkhart \& Lazarian \cite{burk2012}),
    characterizing the
    link between density and velocity in a cloud. We find: \\
  (1) The dispersion $\sigma_{\eta}$ of the PDF for Aquila is 0.3
    and for the two Polaris subregions 0.22 and 0.27, while
  $\sigma_{\eta}$=0.45 for Orion B.  At the same time, the Mach number
  for all regions is typically 6 to 8 (in view of the uncertainty
    of $\mathcal{M}$, they are bascially the same), while only the
  Polaris-quiet subregion has a significant lower value of 3.
  Numerical models indicate that a larger width is caused by a higher
  Mach number and/or compressive forcing instead of solenoidal forcing
  (see also Federrath et al.  \cite{fed2010}, Tremblin et al. 
  \cite{tremblin2012}).  Since Orion B and Aquila have similar values of
  Mach number, we conclude that the Orion B cloud is likely exposed to
  compressive modes -- as seen also in the sharp cutoff of column
  density (Fig.~\ref{orionb}) -- caused by the stellar winds from
  diverse OB aggregates\footnote{Note that the Pipe cloud as a clear
    example of compression (Sec. 3.1) also shows a broad PDF with
    $\sigma_{\eta}$=0.60 in extinction maps (Schneider et al. 2013.)}.
  Aquila has been proposed to be located at an encounter of several
  superbubbles (Frisch \cite{frisch1998}), but the impact on its
    density structure -- more or less important than close-by OB-stars
    -- can not be inferred. Both clouds are exposed to relatively high
    magnetic fields (Crutcher et al.  \cite{crutcher1999}, Sugitani et
    al. \cite{sugitani2011}), so that the more narrow PDF of Aquila is
    presumably not caused by magnetic fields
    alone.  \\
  (2) Polaris-quiet has a narrow ($\sigma$=0.22), lognormal  
    PDF, the gas is nearly isothermal (see 
    temperature PDF in Fig.~\ref{orionb-temp}), and has a low
    ($\sim$3) Mach-number.  Only isothermal turbulence simulations
    without self-gravity reproduce this shape of the PDF.  We computed
    the forcing-parameter 
    $b = 1/\mathcal{M} \times (\exp{((f\,\sigma_{\eta})^2)-1})^{0.5}$ 
    using an average of 2.5 between 
    solenoidal and compressive forcing (Federrath et al. 2010) for the
    factor $f = \sigma_s/\sigma_{\eta}$ (estimation of the 3D density 
    fluctuation $\sigma_s$ out of the 2D column density fluctuation 
    $\sigma_{\eta}$).  The resulting value of
    $b$=0.2 is lower than what was found by comparing with the purely
    solenoidal driven isothermal MHD simulations of Burkhart \&
    Lazarian \cite{burk2012}). Our data point for Polaris-quiet fits
    on their model (Fig. 3) with b=1/3.  In any case, these results
    show that the Polaris-quiet PDF is consistent with the view that
    the cloud's density distribution is mainly
    governed by solenoidal forcing.\\
%The purely turbulent character of the gas phase in Polaris was also shown in power spectra
%from {\sl Herschel} imaging observations (Miville-Desch\^enes et al.
%2010). \\
% where a power law exponent of $\gamma$=-2.6 was fitted for the
%power spektrum P(k)$\sim$k$^{\gamma}$, corresponding to
%$\alpha$=$\gamma$+1=--1.6 and an inferred SFE of 0\% in Federrath \&
%Klessen (2012). \\ 
  3) Power-law tails in the high-density PDF regime form under
    the presence of self-gravity, but can also be provoked by purely
    non-isothermal turbulence (Passot \& Vazquez-Semadeni
    \cite{passot1998}).  For Orion B and Aquila, gravity most likely
    dominates because a large number of pre- and protostellar dense
    cores and supercritical filaments are present (Andr\'e et al.
    2010, K\"onyves et al., in prep.). The gas is not isothermal
    (Fig~\ref{orionb-temp}), but the temperature does not vary by
    several orders of magnitude either. The excess in the PDF for the
    Polaris-saxophone region is more difficult to interpret. In this
    gravitational and thermally subcritical filament, only one
    candidate pre-stellar core (Ward-Thompson et al. \cite{derek2010},
    Shimoikura et al. \cite{shi2012}) was found. The gas can be
    condidered as isothermal, so that here magnetic fields may play a
    role (the strength is not known), leading to a narrow PDF (e.g.,
    Molina et al.  (2012)), or statistical density fluctuations and
    intermittency due to locally compressive turbulence. These effects
    may also explain the slight excess in the PDF of Polaris-quiet.

\acknowledgments
SPIRE has been developed by a consortium of institutes led
by Cardiff Univ. (UK) and including: Univ. Lethbridge (Canada);
NAOC (China); CEA, LAM (France); IFSI, Univ. Padua (Italy);
IAC (Spain); Stockholm Observatory (Sweden); Imperial College
London, RAL, UCL-MSSL, UKATC, Univ. Sussex (UK); and Caltech,
JPL, NHSC, Univ. Colorado (USA). This development has been
supported by national funding agencies: CSA (Canada); NAOC
(China); CEA, CNES, CNRS (France); ASI (Italy); MCINN (Spain);
SNSB (Sweden); STFC, UKSA (UK); and NASA (USA). We thank 
J.-Ph. Bernard for providing the flux  values from Planck. 
C. Federrath acknowledges the Australian Research Council for a 
Discovery Projects Fellowship (grant No. DP110102191). 
Part of this work was supported by the  ANR-11-BS56-010  
project ``STARFICH''.  


%% in parentheses, after the keyword, but they will not be verified.

%{\it Facilities:} \facility{Nickel}, \facility{HST (STIS)}, \facility{CXO (ASIS)}.



%\appendix

%\section{Appendix material}

\end{document}